%Plain TeX
\font\foot=cmr9
\font\norma=cmr10
\font\summ=cmr9                                                    
\font\nom=cmcsc10

\centerline{\bf A forgotten argument by Gordon uniquely selects Abraham's 
tensor} 
\centerline{\bf as the energy-momentum tensor for the electromagnetic field} 
\centerline{\bf in homogeneous, isotropic matter.}\bigskip 
\centerline{{\nom S.~Antoci} and {\nom L.~Mihich}} 
 \centerline{\sl Dipartimento di Fisica ``A.~Volta''~-~Pavia, Italy} 
\bigskip
\bigskip
\noindent {\bf Riassunto.} {\summ - Vista la situazione attuale del problema 
del tensore elettromagnetico dell'energia nella materia \` e forse utile 
rammentare un argomento dimenticato, dato nel 1923 da W.~Gordon. 
Per un mezzo materiale che a riposo sia omogeneo ed isotropo tale 
argomento permette di ricondurre il problema suddetto a un problema 
analogo, definito nel vuoto della relativit\` a generale (in presenza 
di una metrica effettiva $\gamma_{ik}$ opportunamente determinata). 
Per questo secondo problema la forma della lagrangiana elettromagnetica 
\` e nota, e la determinazione del tensore dell'energia si compie subito, 
perch\` e basta eseguire la derivazione della lagrangiana cos\` \i\space scelta 
rispetto alla metrica vera $g_{ik}$. Si seleziona allora il tensore 
di Abraham come tensore elettromagnetico dell'energia per un mezzo 
che a riposo sia omogeneo ed isotropo.
\bigskip
\bigskip

 \noindent {\bf Summary.} {\summ - Given the present status of the problem of the
  electromagnetic   energy   tensor in matter, there   is perhaps use in
recalling a forgotten argument given in 1923 by W.~Gordon. Let   us 
consider a material medium which is homogeneous and isotropic when 
observed in its rest frame. For such a medium, Gordon's argument 
 allows to reduce the above mentioned problem to an analogous   one, 
 defined in a general relativistic vacuum ({\sl i.e.} in   presence of a 
 suitably determined metric $\gamma_{ik}$). For the latter problem 
   the   form of the Lagrangian is known already, hence the
 determination of the energy tensor is a straightforward matter.
  One just   performs   the 
 Hamiltonian derivative of the Lagrangian chosen in this   way   with
  respect to the true metric $g_{ik}$. Abraham's tensor is thus   selected 
 as the   electromagnetic   energy   tensor   for   a   medium   which   is 
 homogeneous and isotropic in its rest frame.}\bigskip

P.A.C.S.:~{\bf 03.50-z}~-~Classical field theory.

P.A.C.S.:~{\bf 04.20-q}~-~Classical general relativity.\filbreak

 \noindent {\bf 1. Introduction.}\bigskip
\norma

     The  reader  of  old  scientific  literature   is   sometimes 
confronted with findings  that  somewhat  shake  his  belief  that 
physics, at variance with other sorts of human endeavours,  undergo
 a more or less steady progress, maybe interlarded from time  to 
time with Kuhn-like revolutions [1], but always driven by rational 
motivations. He is in fact forced to recognize  the  existence  of 
forgotten papers, whose indisputable  relevance  for  some  reason 
went unnoticed at the time  of  their  publication.  Due  to  some 
``paradigm  shift''  and  to  the  associated,  ingrained  habit  of 
physicists to occupy their minds mainly  with  state  of  the  art 
contributions,  it  is  quite  difficult  that  such  papers   may 
resurface from the oblivion in which their were cast at  the  very 
moment of their appearance.

As far as we could ascertain, this was  the  destiny  of  the 
article  ``Zur  Lichtfortpflanzung  nach  der  Relativit\"atstheorie'' 
published in 1923 by W.~Gordon [2]. It is true that this paper was 
published just before the major paradigm shift that placed quantum 
mechanics at center stage and relegated general relativistic ideas 
to a somewhat lesser r\^ole. It is also true that in 1960 J.L.~Synge 
[3] was well aware of the virtues, for the geometrical  optics  of 
non-dispersive media, of the effective metric introduced by Gordon 
in the above mentioned paper. In that book Synge recalls that,  in 
the limit of geometrical optics, the rays are null geodesics  with 
respect to Gordon's effective metric, but  he  does  not  remember 
that the original finding of  this  property  (although  only  for 
homogeneous, isotropic media) is due to Gordon. About this finding 
he mentions in fact only later authors [4],[5].

However, the main reason why Gordon's paper should have  been 
considered at the time of its  appearance  and  remembered  thereafter 
is, in our opinion, another one and, to  our  knowledge,  it 
has remained completely unnoticed by everybody,  if  one  excludes 
the attentive reviewer  of  Gordon's  article  for  ``Physikalische 
Berichte''~[6]:  that  paper  contains  a  clear-cut  argument  for 
selecting, through  a  clever {\sl  reductio  ad  vacuum},  the 
stress-energy-momentum   tensor   of   the electromagnetic field in 
non-dispersive matter, provided that the latter be homogeneous and 
isotropic when considered in its rest frame.

We have found no track  of  Gordon's  argument  in  the  huge 
literature that has been produced on this controversial  issue  in 
the decades elapsed since then; we  feel  therefore  enticed  into 
recalling it here. We do so not for the historical record, but for 
the very reason that the problem  of  the  energy  tensor  of  the 
macroscopic electromagnetic field in matter, after so many efforts 
both on the theoretical and on the experimental  side,  and  after 
the above mentioned paradigm shift, is still with us, as open  and 
unsolved as it was at the time when Minkowski [7] and Abraham  [8] 
provided their well known, diverging answers.
\bigskip
\bigskip

 \noindent {\bf 2. Field  equations  and  constitutive relations for
 electromagnetism in matter and in vacuo.}\bigskip
\norma
In order to best appreciate the general relativistic argument 
conceived by Gordon one should remind of the position occupied  by 
Maxwell's equations in the structure of space-time.

Remarkably enough, neither a metric nor an affine  connection 
are required for their definition; these equations can be  written 
as soon as the concept of four-dimensional differentiable manifold 
is introduced [9]. In such a manifold one can consider  a  contravariant
skew tensor density
\footnote{\foot$^{1)}$}{Boldface letters are henceforth used 
to indicate densities.}
${\bf H}^{ik}$
\norma , a covariant skew tensor $F_{ik}$, and write the 
``naturally invariant'' equations \footnote{\foot $^{2)}$}{For ease of
 comparison, we conform to the notation and to  the  conventions 
adopted by Gordon in Ref. [2].}
\norma
 
            $${\bf H}^{ik}_{  ,k}={\bf s}^i,\eqno(1)$$
and
           $$F_{[ik,m]}=0.\eqno(2)$$
The vector density ${\bf s}^i$, defined by the left-hand side of  eq.~(1), 
is assumed to represent the electric four-current  density,  while 
the comma  signals  ordinary  differentiation,  and  we  have  set 
$F_{[ik,m]}\equiv{\textstyle{1\over3}}(F_{ik,m}+F_{km,i}+F_{mi,k})$. 
Equations (1) and (2) express Maxwell's equations in general curvilinear  
co-ordinates;  they constitute  two  sets  of  four  equations,   which   
fulfil two identities:

   $${{\bf H}^{ik}_{,k,i}={\bf s}^i_{,i}=0,}\eqno(3)$$
that accounts for the conservation of the  electric  four-current, and

      $${\bf e}^{ikmn}F_{[ik,m],n}=0,\eqno(4)$$
where ${\bf e}^{ikmn}$ is the totally antisymmetric symbol of Ricci 
and  Levi Civita. Therefore  equations (1) and (2) cannot  be  used  for 
determining, in a given co-ordinate  system,  both ${\bf H}^{ik}$
and  $F_{ik}$, which possess together 12 independent  components.
Equations  (1) and (2) need to be complemented with the constitutive
relations of electromagnetism, {\sl i.e.} with a set of six tensor equations of
some sort that allow to uniquely define {\sl e.g.} ${\bf H}^{ik}$
in terms of $F_{ik}$ and of whatever additional  fields  may  be  needed 
for  specifying  the properties of the electromagnetic medium that one
is  considering. In this way the number of independent components of
${\bf H}^{ik}$ and $F_{ik}$ is reduced to 6, and eqs.~(1) and (2) may
suffice for predicting  the evolution of the electromagnetic field (for a  given
four-current density ${\bf s}^i$). 

     When the dependence of ${\bf H}^{ik}$ on $F_{ik}$ is assumed to be
algebraic and linear, the constitutive relations can read [10]

   $${\bf H}^{ik}=\textstyle{1\over2}{\bf X}^{ikmn}F_{mn},\eqno(5)$$
and the electromagnetic properties of the medium are summarized by 
the four-index tensor density ${\bf X}^{ikmn}$, which is skew both in the 
first and in the second pair of indices. A particular  example  of 
the constitutive relations (5) is provided by  the  very  peculiar 
medium that we call vacuum. In this case one writes: 

$${\bf X}^{ikmn}_{vac.}=\sqrt{g}(g^{im}g^{kn}-g^{in}g^{km}),\eqno(6)$$
{\sl i.e.} the constitutive relations entail the metric tensor $g_{ik}$    and 
$g\equiv-det(g_{ik})$ in the way well known from general  relativity.  For 
the vacuum case Maxwell's equations are more usually written as:

     $${\bf F}^{ik}_{  ,k}={\bf s}^i,\eqno(7)$$
$$F_{[ik,m]}=0,\eqno(8)$$
in terms of a skew  tensor $F_{ik}$ and of the associated  contravariant
 tensor density ${\bf F}^{ik}\equiv\sqrt{g}g^{im}g^{kn}F_{mn}$.
\bigskip 
\bigskip

 \noindent {\bf 3. The derivation of the electromagnetic field equations and  of
 the energy tensor for the 

  \noindent ~~~vacuum of general relativity.}\bigskip

It is well known [11]  that, if one defines the electromagnetic field
 in terms of the four-potential $\varphi_i$:

        $$F_{ik}=\varphi_{k,i}-\varphi_{i,k},\eqno(9)$$
then the homogeneous equations (8) are {\sl a priori} satisfied, and the 
inhomogeneous equations (7) can be derived  through  the  Hamilton 
principle, by starting from the Lagrangian density

  $${\bf L}=\textstyle{1\over4}{\bf F}^{ik}F_{ik}-{\bf s}^i\varphi_i.\eqno(10)$$
It is less known that the lame variational method described  above 
can be dropped, that {\sl both} sets of Maxwell equations can be derived 
through  the  Hamilton  principle  from  the  Lagrangian   density 
reported above [12]. In fact, there  is  no  need  to  restrict  {\sl a 
priori} $F_{ik}$   to be the curl of a four-vector  $\varphi_i$. 
One  can  instead start with a general skew tensor $F_{ik}$, that
can always  be  written as the sum of the curl of a potential and
of the dual to the  curl of an ``antipotential'':

$$F_{ik}=\varphi_{k,i}-\varphi_{i,k}
+e_{ik}^{~~mn}(\psi_{n,m}-\psi_{m,n}).\eqno(11)$$
Here $e_{ik}^{~~mn}$  is the tensor obtained from the  totally  antisymmetric 
tensor density ${\bf e}^{ikmn}$ in  the  usual  way.  By  asking  that  the 
variations of $A=\int{\bf L}\,dS$  ($\,dS=\,dx^1\,dx^2\,dx^3\,dx^4$) 
with respect to  $\varphi_i$   and to $\psi_i$ separately vanish one
 immediately obtains eqs. (7) and (8).

The well known form of the energy tensor density ${\bf T}_{ik}$
for  the electromagnetic field {\sl in vacuo} is eventually obtained 
through  the general method [11] inaugurated by Hilbert, i.e. 
by performing the variation of the Lagrangian  density  (10) 
with  respect  to  the metric tensor $g^{ik}$:

$${\bf T}_{ik}\equiv2{\delta{\bf L}\over\delta{g^{ik}}}
={\bf F}_i^{~n}F_{kn}-\textstyle{1\over4}
g_{ik}{\bf F}^{mn}F_{mn}.\eqno(12)$$
If Finzi's variational method is adopted, the  derivation  of  the 
energy tensor must be performed by keeping into account  the  just 
obtained homogeneous field equations, which dictate  that  $F_{ik}$     is 
the curl of a four-vector  $\varphi_i$  and does not contain the metric.
\bigskip
\bigskip

 \noindent {\bf 4. The  constitutive  relations of electromagnetism for an 
     isotropic, homogeneous medium 

  \noindent ~~~in general relativity.}\bigskip

The constitutive relations of electromagnetism for a  linear, 
non-dispersive medium will have in general the form of eq.~(5). We 
notice that, since ${\bf X}^{ikmn}$    is skew both in  the  first  and  in  the 
second pair of indices, its components can be given through a $6\times6$ 
matrix.\footnote{\foot$^{3)}$}{The correspondence between the matrix indices
 and  the  pairs of skew tensor indices is assumed to be as follows:

\  1 $\Leftrightarrow$ 41,
\  2 $\Leftrightarrow$ 42, 
\  3 $\Leftrightarrow$ 43,
\  4 $\Leftrightarrow$ 23,
\  5 $\Leftrightarrow$ 31,
\  6 $\Leftrightarrow$ 12.}
\norma
We shall henceforth deal with the case of a medium  which 
is homogeneous and isotropic, when considered at rest.
     In order to recognize such a medium in general relativity  we 
first transform ${\bf X}^{ikmn}$     to a co-ordinate  system  for  which,  at  a 
given event, $g_{ik}=\eta_{ik}\equiv{diag(1,1,1,-1)}$.  In  the  new  co-ordinate 
system ${\bf X}^{ikmn}$     will in general display off-diagonal  components.  We 
can now perform a Lorentz transformation, that will not change the 
form of the metric at the event under question. Suppose that after 
the Lorentz transformation ${\bf X}^{ikmn}$   (in the matrix rendering) becomes 
diagonal, and that the first three components on the diagonal  are 
equal to, say, $-\epsilon$, while the remaining three read $1/\mu$. 
If this  is the case we have to do (at the considered event) with 
an isotropic medium of dielectric constant $\epsilon$ and of magnetic 
permeability $\mu$. If we apply the same procedure at any event, and  we
find that the matrix can always be put in diagonal form with the same
values for the components as before, the medium is also homogeneous.

The constitutive  relations  for  an  isotropic,  homogeneous 
medium can be written in a simple form, that is due  to  Minkowski 
[7], and can be extended without change to general relativity. Let

    $$u^i={dx^i\over\sqrt{-ds^2}}\eqno(13)$$
be the four-velocity of matter, for which $u_iu^i=-1$.  One  defines 
the four-vectors

  $$F_i=F_{ik}u^k,\quad H_i=H_{ik}u^k,\eqno(14)$$
where $H_{ik}\equiv{(1/\sqrt{g})}g_{ip}g_{kq}{\bf H}^{pq}$ is
the covariant tensor associated with ${\bf H}^{ik}$. Then the above
mentioned constitutive relations simply read

  $$H_i=\epsilon{F_i},\eqno(15)$$
  
  $$u_iF_{km}+u_kF_{mi}+u_mF_{ik}=
  \mu\big(u_iH_{km}+u_kH_{mi}+u_mH_{ik}\big).\eqno(16)$$
As shown by Gordon [2], these eight  equations,  that  entail  two 
identities, are equivalent to the six equations:
 
 $$\mu{H^{ik}}=F^{ik}+(\epsilon\mu-1)(u^iF^k-u^kF^i)\eqno(17)$$
that can be easily cast into the form of eq.~(5). But  the  right-hand 
side of eq.~(17) can be rewritten as

     $$F_{rs}\big\{g^{ir}g^{ks}
  -(\epsilon\mu-1)\big(u^iu^rg^{ks}+u^ku^sg^{ir}\big)\big\}.$$
We can freely add the  term $(\epsilon\mu-1)^2u^iu^ku^ru^s$  within  the
curly brackets, since it will give no contribution,  due  to  the  
antisymmetry of $F_{rs}$. Then eq.~(17) comes to read

$$\mu{H^{ik}}=\big(g^{ir}-(\epsilon\mu-1)u^iu^r\big)
    \big(g^{ks}-(\epsilon\mu-1)u^ku^s\big)F_{rs}.\eqno(18)$$
Let us define the ``effective metric tensor''

$$\gamma^{ik}=g^{ik}-(\epsilon\mu-1)u^iu^k,\eqno(19)$$
whose inverse is

$$\gamma_{ik}=g_{ik}+\big(1-{1\over{\epsilon\mu}}\big)u_iu_k;\eqno(20)$$
then we can bring eq.~(17) to the form:

$$\mu{\bf H}^{ik}=\sqrt{g}\gamma^{ir}\gamma^{ks}F_{rs}.\eqno(21)$$
Since $g\equiv-det(g_{ik})$, we shall pose $\gamma\equiv-det(\gamma_{ik})$. Then
the  ratio $\gamma/g$ shall be an invariant. Its calculation
can be performed in the co-ordinate system in which $u^1=u^2=u^3=0$, and
one finds 

$$\gamma={g\over{\epsilon\mu}},$$
so that eq.~(21) can be rewritten as

$${\bf H}^{ik}=\sqrt{\epsilon\over\mu}
\sqrt{\gamma}\gamma^{ir}\gamma^{ks}F_{rs},\eqno(22)$$
which, apart from  the  constant  factor $\sqrt{\epsilon/\mu}$,
is  just  the constitutive relation of the general relativistic vacuum 
for which $\gamma_{ik}$ acts as metric.

We shall henceforth enclose in  round  brackets  the  indices 
which  are  either  moved  with $\gamma^{ik}$  and $\gamma_{ik}$,  or  
generated  by performing the Hamiltonian derivative with respect to
the  latter tensors; therefore eq.~(22) will be rewritten as

$${\bf H}^{ik}=\sqrt{\epsilon\over\mu}
\sqrt{\gamma}F^{(i)(k)}.\eqno(23)$$
\bigskip
\bigskip

\noindent {\bf 5. The Lagrangian for the electromagnetic field in an isotropic, 
 homogeneous medium and the 

  \noindent ~~~energy tensor derived from it.}\bigskip

     If the metric  field  is $\gamma_{ik}$, according to eq.~(10)  the 
Lagrangian density for the electromagnetic field {\sl in vacuo} reads:

   $${\bf L}=\textstyle{1\over4}\sqrt{\gamma}F^{(i)(k)}F_{ik}
   -{\bf s}^i\varphi_i.\eqno(10')$$
where $F_{ik}$  will be given either by eq.~(9) or by

  $$F_{ik}=\varphi_{k,i}-\varphi_{i,k}+{1\over{\sqrt\gamma}}
  e_{(i)(k)}^{~~~~~~mn}(\psi_{n,m}-\psi_{m,n}).\eqno(11')$$
in compliance with the complete  variational  method  proposed  by 
Finzi. Selecting the Lagrangian density  for  the  electromagnetic 
field in the medium under question is then reduced to a  straightforward 
matter. Gordon [2] writes:
 
 $${\bf L}'=\textstyle{1\over4}\sqrt{\epsilon\over\mu}
 \sqrt{\gamma}F^{(i)(k)}F_{ik}-{\bf s}^i\varphi_i,\eqno(24)$$
where $F_{ik}$ can be presently defined by eq.~(11'). Equating to  zero 
the independent variations of the action $\int{{\bf L}'\,dS}$ with respect to
 $\varphi_i$ and to $\psi_i$ will produce the Maxwell's equations (1) and (2) 
respectively, {\sl a priori} complemented by the constitutive  relations 
(15) and (16).

It is now easy to  derive  the  energy  tensor  for   the 
electromagnetic field by starting from  the  derivation  that  one 
performs {\sl in vacuo}, when the metric field $\gamma_{ik}$ is present. 
In that case, in keeping with the general definition (12), one writes:

  $$\delta{\bf L}\equiv{\textstyle{1\over2}}
  {\bf T}_{(i)(k)}\delta\gamma^{ik},\eqno(12')$$
hence one gets

$${\bf T}_{(i)}^{~~(k)}=\sqrt\gamma\big(F_{ir}F^{(k)(r)}
-\textstyle{1\over4}\delta_i^{~k}F_{rs}F^{(r)(s)}\big),\eqno(25)$$
as already shown in \S\ 3. When the matter under question is present 
one instead writes:

$$\delta{\bf L}'\equiv{\textstyle{1\over2}}
  {\bf T}'_{(i)(k)}\delta\gamma^{ik},\eqno(26)$$
where ${\bf L}'$ is given by eq.~(24), and finds

$${\bf T}_{(i)}^{'~~(k)}=F_{ir}{\bf H}^{kr}
-\textstyle{1\over4}\delta_i^{~k}F_{rs}{\bf H}^{rs},\eqno(27)$$
which is  just  the  general  relativistic  version  of  the  form 
proposed by Minkowski in his fundamental work [7]. We  shall  drop 
henceforth the prime in the expression of the energy tensor, since 
its omission will not lead to confusion.

${\bf T}_{(i)}^{~~(k)}$, however, cannot be the energy tensor density
that  we are seeking, because it is defined with respect to  the 
effective metric  $\gamma_{ik}$, not with respect to the true metric 
$g_{ik}$, the only  one that accounts  for  the structure of space-time
and,  via  the Einstein tensor, for its stress-energy-momentum tensor $T_{ik}$.

In order to find the relation  between ${\bf T}_{ik}$ and 
${\bf T}_{(i)(k)}$ we simply need to express $\delta\gamma^{ik}$
in terms of $\delta g^{ik}$. From  eq.~(13) one derives the variation 
of $u^i$  induced by the variation $\delta g^{mn}$ of  the metric:

 $$\eqalign\delta u^i={\textstyle{1\over2}}u^iu^mu^n\delta g_{mn}
=-{\textstyle{1\over2}}u^iu_mu_n\delta g^{mn},\eqno(28)$$
hence from the definition (19) we obtain:

$$\eqalign\delta\gamma^{ik}=\delta\big\{g^{ik}-(\epsilon\mu-1)u^iu^k\big\}
=\delta g^{ik}+(\epsilon\mu-1)u^iu^ku_mu_n\delta g^{mn}.\eqno(29)$$
Therefore, since ${\bf T}_{ik}\delta g^{ik}={\bf T}_{(i)(k)}\delta\gamma^{ik}$,
we get immediately:

$$\eqalign{\bf T}_{ik}={\bf T}_{(i)(k)}
+(\epsilon\mu-1)u_iu_k{\bf T}_{(m)(n)}u^mu^n.\eqno(30)$$

The mixed components of ${\bf T}_{ik}$   can be  obtained  by  multiplying 
the left-hand side and the second term at the right-hand  side  of 
eq.~(30) by $g^{kq}$, while the first term at the  right-hand  side  is 
multiplied by ${\gamma^{kq}+(\epsilon\mu-1)u^ku^q}$. 
In this way we find

$${\bf T}_i^{~q}={\bf T}_{(i)}^{~~(q)}
+(\epsilon\mu-1)\big\{{\bf T}_{(i)(k)}u^k
+u_i{\bf T}_{(m)(n)}u^mu^n\big\}u^q.\eqno(31)$$
In the co-ordinate system for which at a given event $g_{ik}=\eta_{ik}$ and 
$u^1=u^2=u^3=0,\ u^4=1$, the covariant four-vector within the curly 
brackets has the components ${\bf T}_{(\alpha)(4)},\ 0\  (\alpha=1,2,3)$. 
But under  the circumstances chosen above eq.~(30) says that 
${\bf T}_{(\alpha)(4)}={\bf T}_{\alpha4}$,  hence in a general 
co-ordinate system one can write:

$${\bf T}_i^{~q}={\bf T}_{(i)}^{~~(q)}
+(\epsilon\mu-1)\big\{{\bf T}_{ik}u^k
+u_i{\bf T}_{mn}u^mu^n\big\}u^q,\eqno(32)$$
i.e., according to eq.~(27)

$$T_i^{~k}=F_{ir}H^{kr}-\textstyle{1\over4}
\delta_i^{~k}F_{rs}H^{rs}-(\epsilon\mu-1)\Omega_iu^k,\eqno(33)$$
where Minkowski's ``Ruh-Strahl'' [7]

$$\Omega^i=-\big(T_k^{~i}u^k+u^iT_{mn}u^mu^n\big)\eqno(34)$$
has been introduced. Since $\Omega^iu_i=0$, by substituting (33) into (34) 
one finds

$$\eqalign\Omega^i=F_mH^{im}-F_mH^mu^i=u_kF_m
\big(H^{ik}u^m+H^{km}u^i+H^{mi}u^k\big).\eqno(35)$$
Equations (33) and (35) define the extension to general relativity 
of the energy tensor proposed by  Abraham  [8]  for  the  electrognetic 
field in an isotropic, homogeneous medium.
\bigskip
\bigskip

 \noindent {\bf\ 6. Other arguments.}\bigskip

     The weight of Gordon's argument is better appreciated if  one 
considers also other proposals that have been done  for  selecting 
the energy tensor of the electromagnetic field in matter.

     Among them a relevant position is occupied  by  the  argument 
originally outlined by Scheye [13], fully  developed  by  v.  Laue 
[14], and revisited by M{\o}ller [15]. We have already  mentioned  in 
the Introduction that in a homogeneous, isotropic medium the light 
rays, in the limit of geometrical optics, are null  geodesics  [2] 
with respect to the effective metric $\gamma_{ik}$  given by eq.~(20).  This 
result was originally found by Gordon, and  subsequently  extended 
to a non-dispersive, isotropic medium [3]. The derivation of  this 
result only  entails  Maxwell's  equations  and  the  constitutive 
relations; no knowledge of the energy-momentum tensor is  required 
for obtaining it. The ray four-velocity [2], [3] turns out to be a 
timelike four-vector with respect to the true metric $g_{ik}$, {\sl i.e.} the 
ray can accompany a material particle in its  path,  and  keep  it 
permanently illuminated.

In the celebrated paper  ``Zur  Minkowskischen  Elektrodynamik 
der bewegten K\"orper''  [14]  v.~Laue  considers  a  plane  
electromagnetic wave propagating in a medium whose constitutive relations 
are given by eqs.~(15) and (16). For  that  wave  he  introduces  a 
(three-vector) velocity of radiation $v_{\alpha}\ (\alpha=1,2,3)$,
given  by  the ratio between the  three-vector $cT_{4\alpha}$,  that
defines  the  energy current density of  the  electromagnetic  field,
and  its  energy density $T_{44}$:

$$v_{\alpha}={cT_{4\alpha}\over{T_{44}}},\eqno(36)$$
where $c$ is the speed of light in the vacuum of special relativity.
     Laue then postulates the following  selecting  criterion  for 
the energy tensor of the electromagnetic field: $T_{ik}$   must  be  such 
that the timelike four-vector

 $$w_i=\left({v_{\alpha}\over\sqrt{c^2-v^2}}\ ,
 \ {c\over\sqrt{c^2-v^2}}\right)\eqno(37)$$
exist and coincide with the four-velocity of the ray that directly 
stems  from  Maxwell's  equations.   Laue's   criterion   excludes 
Abraham's tensor from the permitted ones, while it allows all  the 
energy tensors whose fourth row is proportional to the fourth  row 
of Minkowski's tensor.

     How seriously shall we take Laue's criterion? It  depends  on 
one's degree of conviction about the ``physical reality'', {\sl i.e.}, for 
the ingrained positivist, about the observability  of  the  energy 
density and of the energy current density of  the  electromagnetic 
field. One cannot forget the impressive examples provided by  Bopp 
[16] for the  odd  behaviour  exhibited  by  the  Poynting  vector 
already {\sl in  vacuo}.\footnote{\foot $^{4)}$}{For instance, consider   
two  electromagnetic  plane waves  whose wavevectors lay in the same 
plane, and make  a  certain  angle $\alpha$. Assume their polarizations to be 
linear, one laying in  the  plane of the wavevectors, the other one 
orthogonal to it [16]. Then  the Poynting vector exhibits in general 
a quite  perplexing  component normal to the plane of the wavevectors!} 
\norma
Nor  one  can  help  sharing  the  tragicomic embarrass of the 
theoretical physicist in  the  exhilarating  tale written by 
J.L.~Synge [17] and  aptly  entitled  ``On  the  present 
status of the electromagnetic energy-tensor''.\footnote{\foot $^{5)}$}{In
that tale a theoretical physicist is bluntly  confronted   with the
questions posed by simple Simon, a man who farms chicken and wants 
to understand for good why in the incubator built by himself 
with static, mutually orthogonal electric and magnetic fields   his 
poor chicken do not appear to benefit from the theoretically 
predicted electromagnetic energy flow.}
\norma

     Another argument for selecting the energy tensor relies on  a 
{\sl reductio ad vacuum} of a different sort, namely to  the  vacuum  of 
special relativity, supposed to prevail at  a  microscopic  scale. 
This is one of  the  pillars  of  a  far-reaching  program,  whose 
initiator was Lorentz himself [18], and which has been endorsed by 
many adherents [19]-[23]. In particular the issue  of  the  energy 
tensor has been attacked by De Groot and Suttorp in  a  remarkable 
series of papers [24] aiming at  a  clean  derivation,  explicitly 
conforming to the tenets of special relativity. But  they  too  do 
not seem to escape a simple criticism, that one {\sl e.g.} reads off  an 
earlier paper by Ott [25].

Imagine that one wishes to produce the  macroscopic  electromagnetic 
energy tensor by starting from  the  vacuum  tensor  that 
supposedly holds at a microscopic scale, in keeping with  Lorentz' 
idea. One is then confronted with the  task  of  calculating  some 
sort of macroscopic average, like

$$<T_{ik}>=<F_i^{~n}F_{kn}-\textstyle{1\over4}
\eta_{ik}F^{mn}F_{mn}>,\eqno(38)$$
where the brackets are used  to  indicate  that  some  process  of 
macroscopic averaging has been performed. Let us assume  that  the 
microscopically fluctuating electromagnetic field $F_{ik}$ can be split 
into a microscopically smooth average term $\bar F_{ik}\equiv{<F_{ik}>}$, 
and a term $\xi_{ik}$, that oscillates at a microscopic scale:

$$F_{ik}={\bar F_{ik}}+\xi_{ik}.\eqno(39)$$
Then, obviously:

$$<T_{ik}>={\bar F}_i^{~n}{\bar F}_{kn}-\textstyle{1\over4}
\eta_{ik}{\bar F}^{mn}{\bar F}_{mn}+<\xi_i^{~n}\xi_{kn}-\textstyle{1\over4}
\eta_{ik}\xi^{mn}\xi_{mn}>.\eqno(40)$$
The first two terms at the right-hand side form the energy  tensor 
of the macroscopic field ${\bar F}_{ik}$,  whose  constitutive  relations 
are just the ones that hold for the macroscopic vacuum. The evaluation 
of the last term looks instead like a  desperate  task,  since  it 
would  require  a  detailed  knowledge  of  correlations   between 
microscopic  fields  for  which  we  have  nothing   better   than 
hypothetical microscopic models.
     But suppose anyway that this task  can  be  accomplished:  we 
still have to split $<T_{ik}>$ into a part pertaining  to  the  medium, 
and a part pertaining to the  macroscopic  field  that  with  that 
medium is interacting. No wonder,  then,  if  the  most  ingenious 
efforts in this direction, spread  on  a  time  span  that  covers 
several decades, do not seem to converge towards a unique  answer, 
as one gathers from the study of the literature mentioned above.
\bigskip
\bigskip

\noindent {\bf 7. Conclusion.}\bigskip

Given the bleak status in  which  the  issue  of  the  energy 
tensor for the electromagnetic field in matter  finds  itself,\footnote
{\foot $^{6)}$}{Not substantially bettered by the improved
description of microscopic matter that can be achieved through 
quantum mechanics.}
\norma
a little paradigm shift suggests itself: rescuing from oblivion  the 
admittedly  macroscopic,  but   clear-cut   general   relativistic 
argument given by Gordon, that uniquely selects  Abraham's  tensor 
as the electromagnetic energy tensor for a material  medium  which 
is homogeneous and isotropic in its rest frame.\filbreak

 \noindent {\bf References}\bigskip

[1]~~{\nom Kuhn~T.S.}, {\sl The  Structure  of  Scientific  Revolutions}  
(University of Chicago Press, Chicago) 1970.

[2]~~{\nom Gordon,~W.}, Ann. Phys. (Leipzig) {\bf 72}, 421 (1923).

[3]~~{\nom Synge,~J.L.}, {\sl Relativity: the General  Theory} (North-Holland,  
Amsterdam) 1960.

[4]~~{\nom Balasz,~N.L.}, J. Opt. Soc. Am. {\bf 45}, 63 (1955).

[5]~~{\nom Pham Mau Quan}, Arch. Rational Mech. Anal. {\bf 1}, 54  (1957); see also:

~~~~~{\nom Pham Mau Quan}, Comptes Rend. Acad.  Sci. (Paris) {\bf 242}, 465 (1956).

[6]~~{\nom Lanczos,~C.}, Phys. Ber. {\bf 40}, 1519 (1923).

[7]~~{\nom Minkowski,~H.}, G\"ott. Nachr., Math.-phys. Klasse (1908), 53.

[8]~~{\nom Abraham,~M.}, Rend. Circ. Matem. Palermo {\bf 30}, 33 (1910).

[9]~~{\nom Schr\"odinger,~E.}, {\sl Space-Time Structure}  (Cambridge  University 
     Press, Cambridge) 1954.

[10]~{\nom Post,~E.J.}, {\sl Formal  Structure  of  Electromagnetics}  (North-Holland 
Publishing Company,

~~~~~Amsterdam) 1962.

[11]~{\nom Hilbert,~D.}, G\"ott. Nachr., Math.-phys. Klasse (1915), 395.

[12]~{\nom Finzi,~B.}, Rend. Accad. Naz. Lincei {\bf 12}, 378 (1952).

[13]~{\nom Scheye,~A.}, Ann. Phys. (Leipzig) {\bf 30}, 805 (1909).

[14]~{\nom v.~Laue,~M.}, Zeitschr. f. Phys. {\bf128}, 387 (1950).

[15]~{\nom M{\o}ller,~C.}, {\sl The  Theory  of  Relativity}  (Clarendon  Press, 
Oxford) 1972.

[16]~{\nom Bopp,~F.}, Ann. Phys. (Leipzig) {\bf 11}, 35 (1963).

[17]~{\nom Synge,~J.L.}, Hermathena {\bf 117}, 80 (1974).

[18]~{\nom Lorentz,~H.A.}, Enzykl. d. math. Wiss. {\bf 5}-2, 200 (1904).

[19]~{\nom Einstein,~A.} and {\nom Laub,~J.}, Ann. Phys. (Leipzig) {\bf 26}, 
541 (1908).

[20]~{\nom Born,~M.}, Math. Ann. {\bf 68},  526  (1910)  (from  notes  left  by 
     H.~Minkowski).

[21]~{\nom D\"allenbach,~W.}, Ann. Phys. (Leipzig) {\bf 58}, 523 (1919).

[22]~{\nom Fokker,~A.}, Phil. Mag. {\bf 39}, 404 (1920).

[23]~{\nom Sauter,~F.}, Zeitschr. f. Phys. {\bf 126}, 207 (1949).

[24]~{\nom de~Groot,~S.R.} and {\nom Suttorp,~L.G.}, Physica {\bf 37}
, 284, 297 (1967);

~~~~~Physica {\bf 39}, 28, 41, 61, 77, 84 (1968).

[25]~{\nom Ott,~H.}, Ann. Phys. (Leipzig) {\bf 11}, 33 (1952).
\bye